\documentclass[prd,twocolumn,showpacs,nofootinbib,preprintnumbers,amsmath,amssymb]{revtex4}

\usepackage{graphicx}
\usepackage{amssymb}
\usepackage{amsmath}
\usepackage{parskip}
\usepackage{epsfig}
\usepackage{dcolumn}
\usepackage{rotating}

\newcommand{\bea}{\begin{eqnarray}}
\newcommand{\eea}{\end{eqnarray}}
\newcommand{\ba}{\begin{eqnarray}}
\newcommand{\ea}{\end{eqnarray}}
\newcommand{\be}{\begin{equation}}
\newcommand{\ee}{\end{equation}}

\newcommand{\ga}{\gamma}

\newcommand{\Ga}{\Gamma}

\newcommand{\Da}{\Delta}

\newcommand{\e}{\epsilon}
\newcommand{\za}{\zeta}

\newcommand{\oa}{\omega}

\newcommand{\p}{\partial}

\renewcommand{\vec}{\mathbf}

\newcommand{\pv}{\vec{p}}

\newcommand{\cD}{{\cal D}}

\newcommand{\cL}{{\cal L}}
\newcommand{\2}{\frac{1}{2}}

\newcommand{\ra}{\rightarrow}
\newcommand{\Ra}{\Rightarrow}

\newcommand{\LF}{\left(}
\newcommand{\RF}{\right)}
\newcommand{\LT}{\left[}
\newcommand{\RT}{\right]}

\newcommand{\narx}{\textit{Preprint} }

\def\rme{\text{e}}
\def\rmd{\text{d}}
\def\rmi{\text{i}}

\begin{document}

\title{A Theory of Neutrino Oscillations and Late Time Acceleration} 
\author{Stephon H.S. Alexander}
\affiliation{Department of Physics and Astronomy, Haverford College
Haverford, PA 19041, USA}

\date{\today}

\begin{abstract}
We provide a microphysical model that connects neutrino oscillations to dark energy, which has predictions of Lorentz and CPT violating neutrino oscillations.  We argue that the DE is a BCS condensate of flavored neutrinos.  As neutrinos propigate in their own condensate they naturally have oscillations proportional to the DE energy density.  All that is assumed in this model is a covariant coupling of neutrinos to gravity and a finite number density of neutrinos in the cosmic rest frame; this situation yields an attractive channel for the formation of a spin zero neutrino condensate leading to late time acceleration self consistently.  Moreover, the vacuum oscillation has two predictions: (1) CPT violating oscillations proportional to the DE density (2) A connection between the evolution of the equation of state of the DE condensate and the neutrino oscillations.  These predictions can be probed independently with future Supernovae and Neutrino Telescopes.


\end{abstract}

\maketitle

\section{Introduction}\label{intro}

While we know observationally that dark energy (DE) makes the universe accelerate on large scales and might require a modification of General Relativity in the IR, nothing is known about how it may interact with matter.  Rectently Ando, Kamionkowski and Mocioiu (AKM) proposed a new way to account for DE-induced Lorentz/CPT violation---introducing neutrino oscillations that are energy independent\cite{Marc}.  Moreover, ever since the discovery of DE a number of coincidences have been identified\cite{Nelson}.  A peculiar coincidence exists between the observed neutrino mass differences between propigation states $\Delta m_{atm}^{2} \sim 10^{-3} eV^{2} $ and the DE density $\rho_{DE} \sim (10^{-3} eV)^{4}$.  Some authors have exploited this fact by seeking to make connections between the evolution of the neutrino density and the quintessence scalar, which leads to promising phenomenology\cite{Nelson,Kaplan:2004dq}.  However, in all models that connects neutrino physics to dark energy, a quesitons still remains: 

\begin{itemize}
\item{Does neutrino oscillations and the emergence of DE come from the same underlying physical phenomenon?}
\end{itemize}

In this paper we attempt to answer the above question by providing a mechanism that non-trivialy ties in neutrino oscillations with the emergence of dark energy.  The idea is straightforward:  Due to a generic four fermion coupling from general relativity a flavored condensate of neutrinos form a scalar bound state (Cooper-pair) in the early universe.  At late cosmological times, the coupled equations of the neutrinos and the scale factor generically drives late time acceleration.  Likewise, the effective action of the condensate interacting with Dirac neutrinos predict two types of neutrino oscillations:

\begin{itemize}
\item{Mass oscillations without the need of a see saw mechanism\footnote{This mechanism assumes no Majorana mass terms and a vanishing coupling constants of the Higgs to Dirac neutrino mass terms.}, that arises from interaction between neutrinos and the dark energy condensate.}
\item{ Lorentz and CPT violating oscillations that are independent of energy in contrast to a $E^{-1}$ decay for mass oscillations. }
 \end{itemize}  
 
The paper is organized as follows: In section II we derive the four-fermi interaction from general relativity coupled to free fermions followed by the evaluation of the effective potential of the neutrino condensate.  Section III gives a description of the mechanism of flavor oscillations.  We conclude with a discussion of precision electroweak tests of the model and future research.

\section{Field Equations}
Our starting point is to discuss how a gravitationally induced neutrino BCS condensate forms in the early universe; due a weak attractive four-fermion interaction and a Fermi surface.   We assume that the early universe is filled with a finite number density of neutrinos and follow the dynamical consequences.  We will find that conditions for condensation arise naturally in general relativity.   In fact general covariance plays an important role in the emergence of the four fermion interaction.  A free fermion field subject to a covariant coupling with the gravitational field induces non-propagating torsion.  When we solve for torsion field we naturally get a four-fermion contact interaction induced in the Einstein-Hilbert action.  Due to the symmetries of FRW space-time a self-consistent Cooper-pairing between fermions of opposite spin is naturally induced. 

To address exactly how the four-fermion interaction arises we must  introduce two independent  fields to gravity: the tetrad, $e_{\mu}^I$, an orthonormal coordinate basis for each point on the manifold, 
 and a spin connection $\omega_{\mu\nu}{}^J$ which connects (parallel transports) the tangent spaces at  different points of the manifold. Note that lower-case Latin letters starting from $\mu,\nu, \dots$ denote spacetime indices,  while capital Latin letters starting from $I,J,\dots$ denote internal Lorentz indices. One can then associate a  $4$-dimensional spacetime metric $g_{\mu\nu}$ via
\be 
g_{\mu\nu} = e_{\mu}^I e_{\nu}^J \eta_{IJ}, 
\ee 
where the Minkowski metric is to be viewed as the metric of the internal space. Internal indices are raised and lowered with the Minkowski metric, while spacetime indices are raised and lowered with the spacetime metric. The requirement that the spin connection be torsion free is simply $\omega_{[\mu\nu]}^{\alpha} = 0$. Let us now rewrite the Einstein-Hilbert action in terms of the spin connection and the vierbein.  

The Einstein-Hilbert action is given by the well-known expression:
\be \label{S_E}
 S_{E} =\frac{M_{pl}^2}{2}\int d^4x det(e)\left[R(\omega,e) - \frac{1}{2} \left(i \bar{\psi} \gamma^{I} e_{I}^{\mu} {\cal{D}}_{\mu} \psi + {\textrm{c.c.}} \right)\right]
\ee
Note that a tetrad based formalism is essential for the inclusion of fermions in the theory, since Dirac spinors live naturally in $SU(2)$. Therefore, covariant derivatives associated with the Dirac action are not the usual $SO(3,1)$ covariant derivatives, but instead are given by  ${\cal{D}}_{\mu} \psi := \partial_{\mu} \psi - \tfrac{1}{4} \omega_{\mu}^{IJ} \gamma_I \gamma_J \; \psi,$ where $\psi$ is a Dirac spinor, and $\gamma^I$ are $4\times 4$ gamma matrices. 

In the absence of fermions, the field equations for the connection gives rise to the metric compatibility (or zero torsion) condition: \be D_{[\mu}e_{\nu]}^a=0 \label{compatibility} \ee which gives rise to the Christoffel symbols and the Einstein Hilbert action.  However, things dramatically change when one includes the covariant coupling of free fermions to GR in the presence of first order GR.  The interaction between the fermions and the connection changes the metric compatibility condition and the equation for the connection.  However, we can solve for the connection and metric compatibility by decomposing the connection into symmetric and antisymmetric pieces:
 \be \omega_{\mu}^{IJ}=\omega^{sIJ}_{\mu}+C_{\mu}^{IJ} \ee
where $\omega^{sIJ} _{\mu}$ is the torsion-free spin
connection satisfying the compatibility condition (\ref{compatibility}) and $C_{\mu}^{IJ}$ is the so called ``contorsion'' tensor. The idea is to integrate out the the contorsion tensor which then will lead us to the more familiar second order formulation of gravity where connections are just the metric dependent Christoffel symbols. This can be achieved simply by imposing the structure equations obtained by varying the action with respect to the connection. Using the identity 
\be \gamma^{I} \gamma^{[J} \gamma^{K]} = - i \epsilon^{IJKL} \gamma_{5} \gamma_{L} + 2 \eta^{I[J} \gamma^{K]}, \ee 
we can express the contorsion tensor in terms of the axial fermion current $J_5^I := \bar{\psi} \gamma_5 \gamma^I \psi,$
 \be  e_I^{\mu} C_{\mu JK}=4\pi G {\ga^2\over \ga^2+1}\LF \2 \e_{IJKL}J_{5}^{L}-{1\over \ga}\eta_{I[J} J_{5K]}\RF .\label{contorsion} \ee
From the above expression for the contorsion tensor it is clear that $C_{mJK}$ is a non-propagating field, its field equations do not have any derivatives on it. Thus ``integrating it out'' is not only equivalent to reinserting its expression (\ref{contorsion}) in the full action classically, but also quantum mechanically. Thus the four-fermion contact interaction term that we are  going to generate in going from the first to second order formalism is quantum mechanically an exact result. This is a  key difference from the ``effective'' contact interaction that one obtains in non-abelian gauge theory where the mediating gauge fields do indeed propagate, and therefore the contact term is only a low energy approximation.

Substituting (\ref{contorsion}) into Holst's action we  find that the action can be written as~\cite{perez} \be
 S = S_{E}[\omega] +  S_{D}[\omega] + S_{int}. \label{split-action}\ee

The first and the second terms are the standard Einstein-Hilbert and Dirac actions involving Christoffel connections. Crucially however, one obtains a  third interaction term given by:
\begin{align} \label{inter-term} 
 S_{int} &=  \frac{3}{2}\pi G \LF{\ga^2\over \ga^2+1}\RF\int d^{4}x \; e \;J_{5I}
J^I_5 \nonumber\\
&\equiv \frac{1}{M_{Pl}^{2}}\int d^{4}x \; e \;\bar{\psi} \gamma_{5} \gamma^{I} \psi \bar{\psi} \gamma_5 \gamma_I \psi 
\end{align}
 Such four-fermion interactions were already observed in Einstein-Cartan theory $\left(\ga^2\ra\infty \text{ limit in (\ref{inter-term})}\right)$, although they are suppressed by a power of Newton's constant (a factor of $1/\kappa$ here).  To form a condensate of fermions the Planck suppression must be transcended, which happens when the universe contracts to Planck densities, as we shall discover in the next section. 

\subsection{Neutrino Superfluiditiy}
The key point in the last section was to understand how covariance leads to the four-fermion interaction when fermions interact with gravity.  Various authors have realized that a weak and attractive interaction leads to a cosmological BCS-like condensation of the fermions~\cite{ABC,Giacosa:2008rw}.   The two known regimes where a  four-fermion channel exists are in the electroweak sector and in gravity.  However, the electroweak case  is repulsive for neutrinos \protect\footnote{For a counterargument using modern arguments in fermi-liquid theory see \cite{Mcelrath}}.  On the other hand the gravitational case does generically allow an attractive interaction and condensation is possible.  In what follows we sketch how one obtains the ``gap equation.''  The four fermion interaction can be replaced with an auxiliary field, rendering the path integration as a gaussian over the fermionic field.  This allows one to explicitly evaluate the path integral involving the fermion self-interactions, leading to an effective potential for the condensate.   The first step is to self-consistently find the generating functional for the condition that the auxiliary field $\Delta = \langle \psi^{\dagger}\psi\rangle$.  After expressing the spinors in the Weyl basis and introducing the auxiliary field, the four-fermion generating functional becomes:
\be
Z=\int [\cD\Da][\cD\xi][\cD\za]\rme^{\rmi(S_{\rm fer}+S_{\rm tree})}
\equiv \int [\cD\Da]\rme^{\rmi S_{\rm eff}}\approx \rme^{\rmi S_{\rm eff}}\big|_{\rm SP}
\ee
where we have integrated the Grassmann fields, defined the effective action $S_{\rm eff}$ (often referred to as $\Ga$ in quantum field theory literature) and approximated the functional integral by the saddle point (mean-field approximation, \cite{Sch99}). The effective action $S_{\rm eff}$ can be evaluated by performing the Gaussian integrals in terms of fermionic coordinates. As usual, one ends up with a fermionic determinant which allows us to express the effective action (see \textit{e.g.} \cite{Sch99,KoS}):
\be
S_{\rm eff}=S_{\rm tree}-\rmi\int \frac{\rmd^4p}{(2\pi)^4} \ln (\det A_p)\,,
\ee
where the tree action is obtained from the interaction action
\bea
S_{\rm int}&=&\int \rmd^4x\, e\LT \frac{(\bar{\psi}\psi)^2}{M^2}\RT \nonumber\\
&=&\int \rmd^4x\, e\LT (\bar{\psi}\psi)\Da-\frac{M^2}{4} \Da^2\RT \nonumber\\
&\equiv& S_{\rm mass}+S_{\rm tree}\,.
\eea
The determinant of $A_p$ is straightforward to compute:
\be
\det A_p=[\oa^2-(|\pv|+\mu)^2-\Da^2][\oa^2-(|\pv|-\mu)^2-\Da^2]\,,
\ee
where $\Delta$ is the auxiliary field at the saddle point. 

The chemical potential is responsible for aligning the neutrinos in the cosmic rest frame and is the underlying reason for the Lorentz violation.  Accordingly, we are left to compute the effective potential
\be
V_{\rm eff}\equiv -\cL_{\rm eff}=\frac{M^2}{4}\Da^2-I\,,
\ee
where
\be
I = \int \frac{\rmd^3\vec{p}}{(2\pi)^3}\ \left[\sqrt{(|\pv|+\mu)^2+\Da^2}+\sqrt{(|\pv|-\mu)^2+\Da^2}\right].
\ee
Plugging $I$ into $V_{\rm eff}$ above leads to
\bea \label{Veff}
V_{\rm eff}&=&\frac{M^2}{4}\Da^2-\frac{\Da^2}{4\pi^2} \LT \frac{\Da^2}{4}\LF N+\2+\ln\Da^2 \RF  \qquad\right. \nonumber\\
&&\left.\qquad\qquad\qquad\,\,\,\,\,-\mu^2\LF N+1+\ln\Da^2\RF \RT.
\eea
The above potential has a minimum given by the \emph{gap equation}

\bea
\frac{\p V_{\rm eff}}{\p \Da}=0 \label{gap}\\
\Ra&M^2=\frac{1}{2\pi^2}&\LT\Da^2(N+1)-2\mu^2(N+2)\right. \nonumber \\
&&\left.\qquad\quad+\LF\Da^2 -2\mu^2\RF\ln\Da^2 \RT. \nonumber
\eea

The potential energy also includes a contribution from the chemical potential as well.  We can obtain the chemical potential from noting that the total number $n_0$ of fermions is given by~\cite{Sch99}
\begin{align} \label{neq}
n_0&=\int \rmd^4 x \ e\bar{\psi}\ga^0\psi=\frac{\delta S_{\rm fer}}{\delta \mu} =-a^3 \frac{\p V_{\rm eff}}{\p \mu} \nonumber\\
&=-a^3\frac{\Da^2\mu}{2\pi^2}\LF N+1+\ln\Da^2\RF\equiv a^3 n\,.
\end{align}
When the system relaxes at the minimum of its potential, the total gap energy density of the fluid is given by
\bea
\rho_{\rm gap}&=&V_{\rm min}+\mu n \nonumber\\
&=& \frac{\Da^2}{32\pi^2}\LF\Da^2-8\mu^2\RF \LF 2N+3+2\ln\Da^2\RF\,\,\,\,\,
\eea

The gap field $\Delta$ is a time dependent function and redshifts as $\sim a^{-3}$, however at late times it tends to a constant, which is why it exhibits a late time de Sitter state.

Solving the EH action (\ref{S_E}) in an FRW background leads to the modified Friedman equations 
\be\label{freq}
H^2=\frac{8\pi}{3M_{Pl}^2}(\rho_{\rm gap}+\rho_{\rm m})\,,
\ee
where $\rho_{\rm m}$ is an extra matter term and 
\be 
\rho_{\rm gap} \simeq \frac{\Delta^{4}}{32\pi^{2}}(2N+3+ 2ln\Delta^{2}) \,.
\ee

\begin{figure} 
\centering
\includegraphics[scale=0.8]{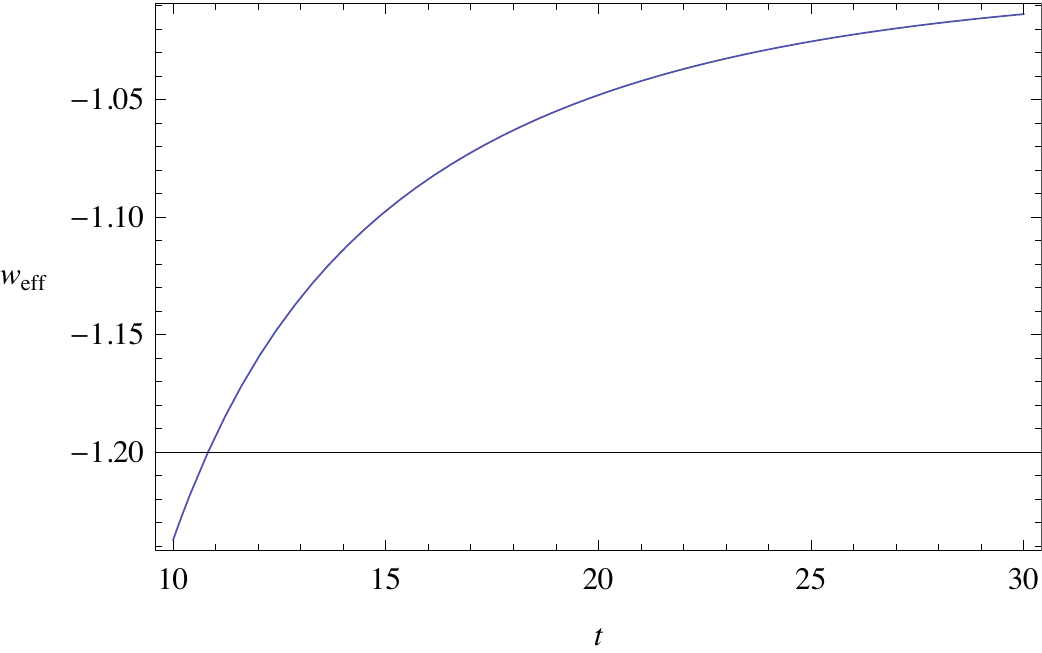}
\caption{This is a plot of the evolution of the dark energy equation of state parameter, $\omega(t)$ of the neutrino condensate.  We see an early phase of Phantom evolution followed by a late time asymptote to a de Sitter phase, ie. $\omega =-1$.}
\label{wplot}
\end{figure}

The cosmological evolution is represented as a transcendental equation that requires numerical treatment.  The authors \cite{ABC} were able to find that the modified Friedmann equation leads to a late time attractor exhibiting acceleration (\textit{i.e.} $\epsilon = -\frac{\dot{H}}{H^{2}} <1$) that depends on the magnitude of the gap.  In particular, when the gap has a mass $M \sim 10^{-3}$  the gap is also mill-electron-volt scale, $\Delta_{0} \sim 10^{-3},$ which corresponds to an energy density with negative equation of state $\omega \simeq -1$.  This condition leads to late time acceleration.  In particular we can solve for the equation of state from the  Raychaudhuri equation:

\be 
\omega(t)_{\rm eff} = -\frac{ \ln\rho(\Delta(t))_{\rm gap}}{\ln{a(t)}} - 1 \, ,
\ee
which is plotted in fig [\ref{wplot}].  Hence, we have established that the neutrino condensate exhibits late time acceleration.
\section{The Mechanism}
We would like to explore how the condensate that results from the four fermion self-interaction induces oscillations.  The key point is that the specific form of the four fermion operator in the presence of flavors naturally mixes the flavor indices due to the Fierz identity.  When the neutrinos of different flavor condensed we are lead naturally to flavor oscillations.  First we need to extend the fermion interactions to (at least 2) flavors.  It is possible to solve (\ref{compatibility}) for N-flavors of fermions:
\be 
 e_I^{\mu} C_{\mu JK}=\sum_{B} 4\pi G {\ga^2\over \ga^2+1}\LF \2 \e_{IJKL}J_{5B}^{L}-{1\over \ga}\eta_{I[J} J_{B5K]}\RF .\label{contorsion} 
\ee
where $B$ is a flavor index.  This leads to a modified interaction term,

\be \cal{L}\rm_{int} = \ \frac{1}{M_{Pl}^{2}}\int d^{4}x \; e \;\bar{\psi}_{A} \gamma_{5} \gamma^{I} \psi_{A} \bar{\psi}_{B} \gamma_5 \gamma_I \psi _{B} \ee

Flavors will get reshuffled due to a Fierz transformation:

\ba  \cal{O}\rm_{\psi} =  \label{flavor} \bar{\psi}_{A} \gamma_{5} \gamma^{I} \psi_{A} \bar{\psi}_{B} \gamma_5 \gamma_I \psi _{B} = \bar{\psi}_{A} \psi_{B} \bar{\psi}_{A} \psi_{B} \nonumber\\ + \bar{\psi}_{A} \gamma_{5} \psi_{B} \bar{\psi}_{A} \gamma_5 \psi_{B}  +  \bar{\psi}_{A}\gamma^{I} \psi_{B} \bar{\psi}_{A} \gamma_I \psi_{B} \ea

 combining  (\ref{inter-term} ) with (\ref{flavor}) and identifying the Dirac spinor with a neutrino field
 $\psi = \nu$
\be 
{\cal{L}}\rm = i\bar{\nu}_{A}\partial_{\mu}\gamma^{\mu}\nu_{A} + \bar{\nu}_{A}\gamma^{\mu} \nu_{B}\Delta_{\mu AB} + (1+ \gamma_{5})\Delta_{AB}\bar{\nu_{A}}\nu_{B} .
\ee

The second term is a Lorentz/CPT violating interaction between the DE condensate and the neutrino.  The third term represents the DE-condensate acting like a Parity and Lorentz invariant masses, $(1+\gamma_{5})\Delta = m_{\Delta}$ respectively.  

To preserve isotropy and homogeneity, the gap must evolve in a frame aligned with the the cosmic expansion, hence the four-vector $\Delta_{\mu} =(\Delta,0,0,0)$.
de Gouvea encountered a similar Lagrangian and interpreted  as a ``matter potential", because it resembles electron neutrinos propagating in the presence of electrons where  $\Delta$ is analogous to the electron number density as neutrinos pass through the sun, \textit{i.e.} \mbox{$\langle e^{\dagger}\sigma_{\mu}e\rangle =(\sqrt{2}G_{F}N_{e},0,0,0)$, where $N_{e}$} is the electron number density \cite{DeGouvea:2002xp}.  However our case represents different physics since the neutrinos are moving in the background of their own BCS-like condensate.   From \ref{condensate} we get the Dirac equations for a neutrino created from the condensate:
\ba \label{diraceq}
(i\partial_{\mu} - \lambda_{AB}\Delta_{\mu})[\bar{\sigma}^{\mu}]^{\dot{a}a}\nu_{aB} - m_{\Delta}^{*}\bar{\nu}_{B}^{\dot{a}} = 0\,, 
\nonumber\\
(i\partial_{\mu} + \lambda_{AB}\Delta_{\mu})[\bar{\sigma}^{\mu}]_{\dot{a}a}\bar{\nu}^{\dot{aB}} - m_{\Delta}\nu_{aB} = 0\,, 
\ea
where $\nu$ is a left handed  neutrino field, $\dot{a}, a=1,2$ are spinor indices and $A, B$ are the flavor indices.  The above Dirac equation has, up to an extra $\gamma_{5}$, the same form as the CPT violating case considered by \cite{Kostelecky:2004hg}.  We now proceed to study the flavor oscillations in the presence of the condensate.

\section{Neutrino Oscillations}
We would like to understand how the presence of the neutrino condensate affects Majorana neutrino oscillations.  For simplicity we will consider a two-flavor oscillation.  The key idea is that the neutrinos of differing flavors can condense.  A neutrino moving in this flavored condensate will experience an oscillation since the condensate has non vanishing off-diagonal components in the flavor eigenbasis.  This analysis dovetails the MSW effect where the equation of motion of neutrinos in a background electron density, $N_{e}$, are taken into account.   Recall that a neutrino in a flavor eigenstate can be decomposed in terms of its mass eigenstates $|\nu_{1}\rangle$ and $|\nu_{2}\rangle$.  For concreteness consider an electron neutrino $|\nu_{e}\rangle$
\be 
|\nu_{e}\rangle = cos\theta|\nu_{1}\rangle + sin\theta|\nu_{2}\rangle \, ,
\ee
where $\theta$ is the mixing angle that parameterizes the mixing matrix $U$.  In general a flavor eigenstate is related to the mass eigenstate by
\be 
|\nu\rangle_{A} = U_{\alpha i}|\nu\rangle_{i} \, ,
\ee 
where \be U_{A i} = \begin{pmatrix}
cos\theta & sin\theta \\
-sin\theta & cos\theta 
\end{pmatrix} \, .
\ee
If a neutrino propagates as a plane wave, then its time development is 
\be 
 \nu(t,x)_{e} = cos\theta e^{-ip_{1\mu}x^{\mu}} \nu_{1}
 + sin\theta e^{-ip_{2\mu}x^{\mu}}\nu_{2} \, .  
 \ee

An ultrarelativistic neutrino of mass $m$, traveling a distance $L$ along the z-direction will have a phase factor $p_{\mu}x^{\mu} = Et-\vec{p}\vec{x}\simeq (E - p_{z})L$.  
We can simplify matters by realizing that, \mbox{$E-p = \frac{(E^{2} - p^{2})}{E + p} \simeq \frac{m^{2}}{2E}$} and $E \simeq |p|$,  leading to: 
\be  
\nu(t,x)_{e} = cos\theta e^{im_{1}^{2}L/2E} \nu_{1} + sin\theta e^{-im_{2}^{2}L/2E} \nu_{2} \, .
\ee  
Now we are ready to describe the time development for neutrinos propagating in the condensate.

We can transform from the mass into the flavor basis by making use of the that fact that $U^{\dagger}U=1$ and multiplying both sides of \ref{aether} by $U_{B i}.$  
In the ultrarelativistic regime, where the neutrino is traveling a distance $L$ and the condensate is purely time-like, the equation of motion (\ref{diraceq}) becomes:
\bea
i\frac{\rm d}{{\rm d}t} 
\begin{pmatrix}
\nu_{e} \\ \nu_{x} 
\end{pmatrix} &=& \left[\frac{\Delta m_{\Delta}^{2}}{2E_{\nu}} \right.
\begin{pmatrix}
sin^{2}\theta  & cos\theta sin\theta  \\ cos\theta sin\theta & cos^{2}\theta 
\end{pmatrix}   \label{de_2} \\
&& \,\,\,\,+ 
\left. \begin{pmatrix}
(\Delta + \sqrt{2}G_F N_e)  & \Delta_{ex}/2 \\ 
\Delta_{ex}^*/2 & 0 
\end{pmatrix} \right]
\begin{pmatrix}
\nu_{e} \\ \nu_{x} 
\end{pmatrix}.\nonumber
\eea
Note the the electron number density $N_{e}$ is included to take into account for the MSW effect.  
 $\nu_x$ is a linear combination of $\nu_{\mu}$ \& $\nu_{\tau},$ and $\Delta\equiv \Delta_{ee}-\Delta_{xx}.$ 
The equation for \mbox{antineutrinos} is identical to Eq.~(\ref{de_2}) with $\Delta_{AB}\rightarrow -\Delta_{AB}$ and 
$N_e\rightarrow -N_e.$  One important difference between the Lorentz preserving mass term and the Lorentz violating coupling is
 a factor $E^{-1}$ in the mass coupling.
 
This was pointed out in \cite{Marc} and can have observational consequences at ultra high energies.  In the absence of matter effects the oscillation probability is 
\be
P_{ex}=P_{xe}\sin^22\theta_{\rm eff}\sin^2\left(\frac{\Delta_{\rm eff}}{2}L\right)
\label{pex}
\ee
where
\begin{align}
&\Delta_{\rm eff}=\sqrt{(\tilde{\Delta}sin2\theta+\Delta_{ex})^2+(\tilde{\Delta}\cos2\theta-\Delta)^2},\\
&\Delta_{\rm eff}\cos2\theta_{\rm eff}=\tilde{\Delta}\cos2\theta-\Delta, \,\\
&\Delta_{\rm eff}\sin2\theta_{\rm eff}=\tilde{\Delta}\sin2\theta+\Delta_{ex} , \, 
\end{align}
and $\tilde{\Delta}\equiv \Delta m_{\Delta}^2/2E_{\nu}$. Henceforth $\Delta_{ex}$ is assumed to be real.  We immediately see that even when we set all masses $m$ to zero, there will still be oscillations in terms of the dark energy condensate
\be \label{condition} 
\Delta_{\rm eff} =\sqrt{\Delta_{e\mu}^{2} +(\Delta_{ee}- \Delta_{\mu \mu})^{2}} \,.
\ee   

In particular, the equation of state $\omega$ depends on the logarithm of the gap's density
\be 
1 + \omega(t)_{\rm eff} = -\frac{ \ln\rho(\Delta(t))_{\rm gap}}{\ln{a(t)}} \, ,
\ee 
so we can obtain a correlation between the equation of state evolution $\omega_{\rm eff}$ and the oscillation angle.   

Recently AKM performed an analysis of CPT violating oscillations\cite{Marc}.  These oscillations were parametrized by a four vector $a_{\mu}$ which has the same algebraic form as our condensate $\Delta_{\mu}$.  Therefore we can map their parameters for Lorentz violation in to ours and use it to constrain the mass term obtained from the DE interacion $m_{\Delta} =\Delta$.   We find that:
\bea m_{eff} sin2\Theta_{d}e^{i\eta} = \lambda_{AB} \Delta \nonumber \\
m_{eff}sin2\Theta_{d}e^{-i\eta}=\lambda_{AB} \Delta^{*} \eea
where $\lambda_{AB}$ is the coupling constant matrix, defined by $\lambda_{AB}\Delta = \Delta_{AB}$.
AKM found that $m_{eff} \sim \frac{\delta\lambda}{\lambda} a^{0} $.  Where $\delta\lambda$ is difference between the eigenvalues of the coupling matrix.    In terms of our neutrino condensate the effective mass is,
$m_{eff} \sim \delta\lambda \Delta.$  AKM found that future experiments that target cosmogenic ultra-high-energy ($\sim 10^{-17}-10^{-19} GeV$) neutrinos produced by the interaction of ultra-high-energy cosmic rays with CMB photons, can improve the sensitivity to down to $\delta\lambda \Delta \sim 10^{-30} GeV$.  If we assume that $\delta \lambda \sim O(1)$ then the dark energy neutrino condensate amplitude can be constrained from future high-energy neutrino observations. 

  \section{Discussion}
In this note we have considered the effects of general relativity covariantly coupled to neutrino fields.  In an FRW background a gravitational contact attraction drives free neutrinos at finite number density into a BCS-like condensate.  Such a condensate has been shown to exhibit the observed late time acceleration provided that the condensate has a milli-electron-volt VEV.  In this work we showed that at late times if the condensate has off diagonal flavor components neutrinos that propagate on cosmological scales will exhibit flavor oscillations, potentially connecting the scale of neutrino oscillations with the scale of dark energy today.  

One immediate concern is whether or not the repulsive interaction that exists between neutrinos and Z-bosons can prevent the gravitationally induced Cooper-pairing of neutrinos from forming.  Likewise, if the condensate forms in the early universe, competing repulsive interactions and scattering can break apart the pairs.   The condensate has correlations on the order of the Hubble radius at the time that it was formed.   During the electroweak epoch $t_{EW} \sim 10^{-12} s$, the repulsive neutrino self-interaction is mediated by the $Z^{o}$ boson.  The effective 4-point operator is:
\be 
\frac{g^{2}_{Z}}{2M_{Z}^{2}}\int \nu_{A}^{\dagger}\bar{\sigma}^{\mu}\nu_{B}\nu_{A}^{\dagger}\bar{\sigma}_{\mu}\nu_{B} \, .
\ee
The weak bare coupling is clearly larger than the gravitational coupling.  However, the condensation mechanism requires a coupling between neutrinos of different flavors.  The Z-boson only couples neutrinos of the same flavor so it will not compete with the gravitational attraction of neutrinos.    

Furthermore the the temperature of the early universe during the electroweak epoch is $T_{EW} \sim 10^{3} GeV.$  Therefore, the $SU(2)$ vector bosons are in equilibrium with the universe during this time.  As a result the interactions due to W and Z bosons will be screened by the mechanism of Debye screening.  The interaction potential for a SU(N) gauge theory with $N_{f}$ fermions was found to be\cite{kobes}:
 \bea 
 U(r)_{SU(N)} \sim  \frac{e^{-m_{D}r}}{4\pi r}  \nonumber \\ 
 m_{D} =(\frac{N_{c}}{3} + \frac{N_{f}}{6} )^{1/2} G_{f}T \,.
 \eea
 Clearly during the electroweak epoch $U(r,T_{EW} )_{SU(N)} \rightarrow 0$ so the four-fermion electroweak exchange is suppressed in the early universe.  The Debye screening will not affect the gravitational interaction since gravity is out of equilibrium at temperatures below the Planck scale.

Another issue to investigate in the future is the stability of the condensate \footnote{The issue of instability of the condensate is currently being investigated by the author and Niayesh Afshordi.}. This issue is important since similar MaVaN models that couple quintessence to neutrino generically exhibit a linear instability with an imaginary speed of sound.  The onset of the instability is close to the time that neutrinos become non-relativistic.  One expects that since our DE condensate couples to neutrinos in a similar manner to MaVaNs that an instability is inevitable. 

 The speed of sound $c_{s}^{2}$ for an adiabatic perturbation of the condensate is\cite{Afshordi}:
\be \label{speedofsound} c_{s}^{2} = \frac{\dot{P}}{\dot{\rho}} = \omega - \frac{\dot{\omega}}{3H(1+ \omega)} \ee

In a future work we will analytically prove that perturbations of the fluid is stable (ie. the speed of sound is positive).  The speed of sound is always positive (\ref{speedofsound}), as we can see from fig [\ref{wplot}] that the time derivative of $\omega$ is positive and the denominator in the second term is negative (since $\omega <-1$).  Therefore, hurestically we see that the speed of sound for perturbations of the condensate is positive and stable.  

 There are a few things left for future investigation.  First, we would like to understand how this type of oscillation might affect other sectors of the standard model such as flavor changing neutral currents.    In particular, neutrino scattering with the condensate can enhance flavor oscillations.  It will be interesting to evaluate these effects explicitly for future neutrino beam experiments.  
 
\begin{acknowledgments}

I especially thank Marc Kamionkowski for contributing to this work at the beginning stages during my visit to Cal Tech.  I also thank Andre de Gouvea for enlightening discussions.  I also thank BJ Bjorken, Peter Love, Dave Spergel and Giovanni Amelino Camelia for discussions.
This work was supported by an NSF CAREER grant.
\end{acknowledgments}



\begin{thebibliography}{57}
\expandafter\ifx\csname natexlab\endcsname\relax\def\natexlab#1{#1}\fi
\expandafter\ifx\csname bibnamefont\endcsname\relax
  \def\bibnamefont#1{#1}\fi
\expandafter\ifx\csname bibfnamefont\endcsname\relax
  \def\bibfnamefont#1{#1}\fi
\expandafter\ifx\csname citenamefont\endcsname\relax
  \def\citenamefont#1{#1}\fi
\expandafter\ifx\csname url\endcsname\relax
  \def\url#1{\texttt{#1}}\fi
\expandafter\ifx\csname urlprefix\endcsname\relax\def\urlprefix{URL }\fi
\providecommand{\bibinfo}[2]{#2}
\providecommand{\eprint}[2][]{\url{#2}}

\bibitem{Nelson}

  R.~Fardon, A.~E.~Nelson and N.~Weiner,
  JCAP {\bf 0410}, 005 (2004)
  [arXiv:astro-ph/0309800].
\bibitem{Kaplan:2004dq}
  D.~B.~Kaplan, A.~E.~Nelson and N.~Weiner,
  Phys.\ Rev.\ Lett.\  {\bf 93}, 091801 (2004)
  [arXiv:hep-ph/0401099].

\bibitem{Marc}
  S.~Ando, M.~Kamionkowski and I.~Mocioiu,
  arXiv:0910.4391 [hep-ph].
\bibitem{Alexander:2005vb}
  S.~Alexander,
  Phys.\ Lett.\  B {\bf 629}, 53 (2005)
  [arXiv:hep-th/0503146].
\bibitem{Giacosa:2008rw}
  F.~Giacosa, R.~Hofmann and M.~Neubert,
  JHEP {\bf 0802}, 077 (2008)
  [arXiv:0801.0197 [hep-th]].
\bibitem{Sch99} Schakel A M J, 1999 \narx{arXiv:cond-mat/9904092}
\bibitem{KoS}   Kogut J B and Stephanov M A, 2004 {\it Camb.\ Monogr.\ Part.\ Phys.\ Nucl.\ Phys.\ Cosmol.} {\bf 21} 
\bibitem{DeGouvea:2002xp}
  A.~De Gouvea,
  Phys.\ Rev.\  D {\bf 66}, 076005 (2002)
  [arXiv:hep-ph/0204077].
\bibitem{Kostelecky:2004hg}
  V.~A.~Kostelecky and M.~Mewes,
  Phys.\ Rev.\  D {\bf 70}, 076002 (2004)
  [arXiv:hep-ph/0406255].
\bibitem{Sch99} Schakel A M J, 1999 \narx{arXiv:cond-mat/9904092}
\bibitem{KoS}   Kogut J B and Stephanov M A, 2004 {\it Camb.\ Monogr.\ Part.\ Phys.\ Nucl.\ Phys.\ Cosmol.} {\bf 21} 1
\bibitem{ABC}
  S.~Alexander, T.~Biswas and G.~Calcagni,
  arXiv:0906.5161 [astro-ph.CO].

\bibitem{Afshordi}
  N.~Afshordi, M.~Zaldarriaga and K.~Kohri,
  Phys.\ Rev.\  D {\bf 72}, 065024 (2005)
  [arXiv:astro-ph/0506663].
\bibitem{Mcelrath}
  B.~McElrath,
  arXiv:0812.2696 [gr-qc].


\bibitem{kobes}
  R.~Kobes, G.~Kunstatter and A.~Rebhan,
  Phys.\ Rev.\ Lett.\  {\bf 64}, 2992 (1990).

\end{thebibliography}

\end{document}